\documentclass[aps,prd,twocolumn,superscriptaddress,nofootinbib]{revtex4-2}
\usepackage{amsmath}
\usepackage{graphicx}
\allowdisplaybreaks
\usepackage[hidelinks]{hyperref}
\usepackage{color}

\usepackage{abrev-jour-long}

\begin{document}

\title[Chaotic orbital dynamics of pulsating stars]{Chaotic orbital dynamics of pulsating stars around black holes surrounded by dark matter halos}
\author{Tiago S. Amancio}
\email{amancio@ifi.unicamp.br}
\affiliation{Instituto de F\'isica Gleb Wataghin, Universidade Estadual de Campinas, 13083-859, Campinas, S\~ao Paulo, Brazil}
\author{Ricardo A. Mosna}
\email{mosna@unicamp.br}
\affiliation{Instituto de F\'isica Gleb Wataghin, Universidade Estadual de Campinas, 13083-859, Campinas, S\~ao Paulo, Brazil}
\affiliation{Departamento de Matem\'atica Aplicada, Universidade Estadual de Campinas, 13083-859,  Campinas,  S\~ao Paulo,  Brazil}   
\author{Ronaldo S. S. Vieira}
\email{ronaldo.vieira@ufabc.edu.br}
\affiliation{Centro de Ci\^encias Naturais e Humanas, Universidade Federal do ABC, 09210-580 Santo Andr\'e, SP, Brazil}
\date{\today}

\begin{abstract}
We analyze the orbital dynamics of spherical test bodies in ``black hole surrounded by dark matter halo'' spherically symmetric spacetimes. When the test body pulsates periodically (such as a variable star), altering its quadrupole tensor, Melnikov's method shows that its orbital dynamics presents homoclinic chaos near the corresponding unstable circular orbits however small the oscillation amplitude is. Since for supermassive black holes the period of revolution of a star near the innermost stable circular orbit roughly spans time intervals from minutes to hours, the formalism can be applied in principle to the astrophysical scenario of a pulsating (variable) star inspiraling into a supermassive black hole, including the black hole SgrA* at the center of our Galaxy.
The chaotic nature of its orbit, due to pulsation, is imprinted in the redshift time series of the emitted light and can, in principle, be observed in the corresponding light curves and even in gravitational-wave signals detected by future observatories such as the Laser Inteferometer Space Antenna. Also, although periodic with respect to the star's proper time, the chaotic orbital motion will produce an erratic light curve (and gravitational-wave signal) in terms of observed, coordinate time. Although our results were obtained for a specific exact solution, we argue that this phenomenon is generic for pulsating bodies immersed in black hole spacetimes surrounded by self-gravitating fluids.
\end{abstract}


\maketitle

\section{Introduction}

It is a well-established fact that point-particle orbits may exhibit chaotic behavior when subject to an external perturbation (see, for example, \cite{lichtenbergLieberman1992, contopoulosOCDA2002, binneytremaineGD}). Both regular and chaotic motion have been the focus of extensive research across astrophysics \cite{contopoulosOCDA2002, binneytremaineGD}, encompassing a wide range of scales, from planetary dynamics \cite{barrow1997poincare, MurrayDermott2000, lecarEtal2001ARAA, ferrazmeloMichtchenkoEtal2005LNP, vieiraRamosCaro2023NewA} to galactic phenomena \cite{binneytremaineGD, contopoulosOCDA2002, henonheiles1964AJ, contopoulos2003galaxies, grosbol2003LNP, grosbol2002SSRv, hunter2003LNP, hunter2005NYASA, quillen2003AJ, ramoscaroLopezsuspesGonzalez2008MNRAS, zotos2011CSF, pichardoEtal2003ApJ, michtchenkoVieiraBarrosLepine2017AA, lepineVieiraEtal2017ApJ, michtchenkoVieiraEtal2018ApJL, contopoulos1960ZA, contopoulos1963AJ, contopoulos1967rta2, dezeeuw1985MNRAS, dezeeuw1988NYASA, binneyMcmillan2011MNRAS, binney2010MNRAS, binney2012MNRASdynamical, binney2012MNRASactions, binneySanders2014IAUS, bienaymeRobinFamaey2015AA, vieiraRamosCaro2016CeMDA, vieiraRamoscaro2014ApJ, vieiraRamoscaro2019MNRAS, vieiraRamoscaroSaa2016PRD}, and extending to cosmological contexts \cite{barrow1982PhysRep, chernoffBarrowPhysRevLett1983, contopoulosEtal1999, motterLetelier2001PLA}.

Within the framework of general relativity, chaotic dynamics has been explored in various contexts, such as in systems involving a black hole perturbed by a gravitational wave \cite{bombelli1992CQGra, letelierVieira1997CQGra}, elastic string loops surrounding black holes \cite{larsen1994CQG, frolovLarsen1999CQG, basuEtal2011PhLB, stuchlikKolos2012PRD, stuchlikKolos2012JCAP}, or in spacetimes featuring a quadrupole deformation of a central mass \cite{gueronLetelier2001PRE, gueronLetelier2002PRE}. Chaotic trajectories have also been identified around spherical black holes surrounded by axisymmetric structures, particularly when the motion occurs outside the equatorial plane \cite{saaVenegeroles1999PhLA, wuZhang2006ApJ, lukesgerakopoulos2012PhRvD, semerakSukova2010MNRAS, semerakSukova2012MNRAS, sukovaSemerak2013MNRAS, witzanySemerakSukova2015MNRAS, polcarSukovaSemerak2019ApJ}, compact objects with radiation fields acting on the particles \cite{defalcoBorrelli2021PRDkerr, defalcoBorrelli2021PRDtimescales}, and in non-Kerr black hole spacetimes arising from modified gravity theories \cite{apostolatos2009PhRvL, contopoulosEtal2011IJBC, mukherjeeEtal2023PhRvD, destounisKokkotas2023GReGr, lukesgerakopoulosEtal2010PhRvD, zelenkaEtal2020PhRvD, destounisEtal2020PhRvD, destounisEtal2021PhRvL, destounisEtal2021PhRvD, destounisEtal2023GReGr, destounisEtal2023PhRvD}.

Recent studies have highlighted that spinning test particles can also exhibit chaotic motion in a Schwarzschild background \cite{suzukiMaeda1997PRD}. Building on this understanding, we recently investigated the possibility of chaotic dynamics in systems involving extended bodies with zero spin. Our findings reveal that introducing finite-size effects can induce chaos in the test body's orbital dynamics when compared to the otherwise integrable trajectories of test particles, both in Newtonian \cite{vieiraMosna2022CSF} and general relativistic contexts \cite{mosnaRodriguesVieira2022PRD, rodriguesMosnaVieira2024GRG}. These works give rise to a natural question: could these findings have applications to the study of pulsating stars within these frameworks? 

Pulsating stars are common in the Universe, being observed since the 17th century. The linear theory of stellar pulsation is well understood \cite{carroll2006introduction}, explaining periodic oscillations in brightness and size. However, to the best of our knowledge, the effects of these pulsations on the orbital dynamics of these stars have not been considered in the literature.
We will model these objects as spherical test bodies. Since in Schwarzschild spacetime they follow geodesics, as we comment later, the presence of gravitating background matter is required if we want to analyze the deviations from geodesic motion due to finite-size effects generated by the body structure. The way we choose to deal with the ambient background matter ``superposed'' to a Schwarzschild black-hole field is by considering a self-gravitating dark-matter halo around it \cite{cardosoEtal2022PRD, konoplyaZhidenko2022ApJ, xuEtal2018JCAP, nampalliwarEtal2021ApJ, jusufi2023EPJC, shenEtal2024PhLB, datta2024PhRvD, daghigKunstatter2024PhRvD,	yangEtal2024EPJC}, since it is ubiquitous in disk and elliptical galaxies \cite{navarroFrenkWhite1996ApJ, navarroFrenkWhite1997ApJ, sofueRubin2001ARAA, einasto2009arXiv, sanders2010darkmatter} (as for instance surrounding supermassive black holes in galactic centers). We argue, however, that this phenomenon exists for any ambient self-gravitating matter field around the black hole.

In this paper, we consider spherically symmetric metrics of the general form
\begin{equation}
	ds^2=-f(r)\,dt^2+g(r)\,dr^2+r^2\left(d\theta^2+\sin^2\theta\, d\phi^2\right),
\label{genmetric}
\end{equation}
in order to model the gravity of a black hole surrounded by a dark matter halo. The method works for any choice of the functions $f(r)$ and $g(r)$; we employ here both exact and approximate solutions to Einstein's field equations found by \cite{cardosoEtal2022PRD, konoplyaZhidenko2022ApJ}. The stellar pulsation is modeled up to quadrupolar order in Dixon's formalism. We search for chaos in the neighborhood of the unstable homoclinic point-particle orbit via Melnikov's method, and apply the results to the scenario of pulsating stars around black holes surrounded by dark matter halos. We adopt the signature ``$-+++$'' and units such that $G=c=1$.

This work is organized as follows. In Sec.~\ref{dixon}, we present the main aspects of Dixon's formalism, and apply it to an oscillating spherical body in a spacetime representing a black hole surrounded by a galactic dark matter halo. In Sec.~\ref{pointparticle}, we analyze homoclinic point-particle orbits in the same spacetime, which are necessary for the application of Melnikov's method, carried out in Sec.~\ref{chaos}. This method indicates that the spherically symmetric oscillations trigger chaotic behavior in the system. Section~\ref{implications} discusses the validity and possible implications of the obtained results in astrophysical observations. Our conclusions are presented in Sec.~\ref{conclusion}.

\section{Dixon's Formalism for Spherically Symmetric Bodies}
\label{dixon}


Dixon's formalism provides a covariant framework to deal with the dynamics of extended test bodies in the context of general relativity. In a series of three articles \cite{DixonI, DixonII, DixonIII}, Dixon proposed a set of covariant equations of motion for the linear momentum $p^\mu$ and the spin tensor $S^{\mu\nu}$ of the extended body.

Consider an extended body represented by the energy-momentum tensor $T^{\mu\nu}$ in a spacetime $\mathcal{M}$, with a metric tensor $g_{\mu\nu}$. We suppose that the effect of the extended body on the geometry of $\mathcal{M}$ is negligible. The dynamics of the extended body is given by the conservation equations
\begin{equation}
	\nabla_\nu T^{\mu\nu}=0.
\label{conserveq}
\end{equation}

In Dixon's formalism the energy-momentum tensor $T^{\mu\nu}$ is expanded in a set of multipole moments. This is done by choosing a spacetime foliation by spacelike hypersurfaces $\Sigma_s$ relative to the worldline of the center of mass of the body, $z^{\mu}(s)$. The evolution parameter $s$ is arbitrary and the center of mass is defined implicitly by the supplementary condition
\begin{equation}
	S^{\mu\nu}(s)\,p_{\nu}(s) = 0.
\end{equation}

With these choices, Eq.~(\ref{conserveq}) is equivalent to the following set of ordinary differential equations:

\begin{eqnarray}
	\frac{Dp^\mu}{ds}&=&-\frac{1}{2}{R^\mu}_{\nu\alpha\beta}\,v^\nu S^{\alpha\beta}+F^\mu,\label{dpdixon} \\
	\frac{DS^{\mu\nu}}{ds}&=&2p^{\left[\mu\right.}v^{\left.\nu\right]}+N^{\mu\nu},\label{dsdixon}
\end{eqnarray}
where ${R^\mu}_{\nu\alpha\beta}$ is the Riemann tensor and $v^\nu=dz^\nu/ds$ is the tangent vector to the worldline.
The terms $F^\mu$ and $N^{\mu\nu}$ are interpreted as the force and the torque relative to the center of mass, respectively, and their expressions up to quadrupole order are given by
\begin{eqnarray}
	F^\mu&=&-\frac{1}{6}J^{\alpha\beta\gamma\delta}\nabla^\mu R_{\alpha\beta\gamma\delta},\label{forcedixon}\\
	N^{\mu\nu}&=&\frac{4}{3}J^{\alpha\beta\gamma\left[\mu\right.}{R^{\left.\nu\right]}}_{\gamma\alpha\beta},\label{torquedixon}
\end{eqnarray}
where $J^{\alpha\beta\gamma\delta}$ is the quadrupole moment of the body and can be chosen to have the same symmetries of the Riemann tensor:
\begin{equation}
	J^{\alpha\beta\gamma\delta}=J^{\left[\alpha\beta\right]\gamma\delta}=J^{\alpha\beta\left[\gamma\delta\right]},
	\hspace{1cm}
	J^{\left[\alpha\beta\gamma\right]\delta}=0.
\label{Jsym}
\end{equation}
It is important to notice that, in Dixon's formalism, the quadrupole moment $J^{\alpha\beta\gamma\delta}$ is arbitrary, only subject to the usual energy conditions. In other words, any such prescribed values for $J^{\alpha\beta\gamma\delta}$ will be consistent with the theory.

With respect to the spacetime symmetries, the formalism demonstrates that each Killing vector \(\xi\) on the manifold \(\mathcal{M}\) is associated with a conserved quantity:
\begin{equation}
    \mathcal{P}_\xi = p_\mu \xi^\mu + \frac{1}{2} S^{\mu\nu} \nabla_\mu \xi_\nu,
\end{equation}
which holds at all orders of perturbation. For the metric presented in Eq.~(\ref{genmetric}), we identify two independent Killing vector fields, \(\partial_t\) and \(\partial_\phi\). In the case of a spinless extended body, which is the case of interest here, this results in the conservation of \(p_t\) and \(p_\phi\).

We choose as the evolution parameter $s=\tau$, the proper time of an observer moving along the center-of-mass worldline. This gives us the normalization $v_\mu v^\mu=-1$.

\subsection{Spherical extended body}

We concentrate now in the case where both the extended body and the metric tensor $g_{\mu\nu}$ have spherical symmetry. The quadrupole tensor takes a simpler form in a frame $\left\{e_a\right\}$, comoving with the body. This reference frame is constructed in the same way as in \cite{mosnaRodriguesVieira2022PRD}. We perform the calculations in this frame and then express them in the coordinate frame. The spherical symmetry of the body, together with Eq.~(\ref{Jsym}), implies that the only components of the quadrupole tensor that may be nonzero are
\begin{eqnarray}
	j_{m}:=J_{0101}=J_{0202}=J_{0303},\\
	j_s:=J_{2323}=J_{1313}=J_{1212}.
\end{eqnarray}

By the symmetries of the problem, a spinless spherical extended body in the equatorial plane will remain spinless and confined to this plane. This follows also from the equations of motion, Eqs.~(\ref{dpdixon}) and~(\ref{dsdixon}), of course. Hence, in what follows, we take consistently $\theta=\pi/2$ and $S^{\mu\nu}=0$, which simplifies the equations of motion to
\begin{equation}
	\frac{Dp_\mu}{d\tau}=F_\mu.
\label{eqmotionsimp}
\end{equation}
The nonzero components of the force, $F_t$, $F_r$, and $F_\phi$, calculated through Eq.~(\ref{forcedixon}) are shown in Appendix~\ref{appendix}, Eqs.~(\ref{ft})--(\ref{fphi}).

Still according to the procedure of \cite{mosnaRodriguesVieira2022PRD}, Eq.~(\ref{eqmotionsimp}) can be solved for the velocities $\dot{t}$, $\dot{r}$, and $\dot{\phi}$, (where the dot means derivation with respect to $\tau$), to obtain a set of equations of the form
\begin{eqnarray}
	\dot{t}=h_1(r,p_r,\dot{p}_r,\tau),\label{velt}\\
	\dot{r}=h_2(r,p_r,\dot{p}_r,\tau),\label{velr}\\
	\dot{\phi}=h_3(r,p_r,\dot{p}_r,\tau)\label{velphi},
\end{eqnarray}
where the conserved quantities $p_t$ and $p_\phi$ only appear as fixed parameters. The dependence on $\tau$ arises from the components of $J_{\alpha\beta\gamma\delta}$ only. These equations can be substituted in $v_\mu v^\mu=-1$ to get an equation of the form $\dot{p}_r=h_4(r,p_r,\tau)$. This expression can be used to eliminate $\dot{p}_r$ from Eq.~(\ref{velr}), resulting in the following set of equations:
\begin{eqnarray}
	\dot{p}_r=h_4(r,p_r,\tau),\label{h4}\\
	\dot{r}=h_5(r,p_r,\tau),\label{h5}
\end{eqnarray}
a nonautonomous dynamical system describing the radial translational motion of the extended body.

When expanded up to quadrupolar order, Eqs.~(\ref{h4}) and~(\ref{h5}) lead to
\begin{eqnarray}
	\dot{r}= f_1(r,p_r)+j_m(\tau)\,g_{1m}(r,p_r)+j_s(\tau)\,g_{1s}(r,p_r),\label{rdot}\\
	\dot{p}_r=f_2(r,p_r)+j_m(\tau)\,g_{2m}(r,p_r)+j_s(\tau)\,g_{2s}(r,p_r).\label{prdot}
\end{eqnarray}
The functions introduced above are shown explicitly in the Appendix, where we have defined $E=-p_t$ and $L=p_\phi$. Also, it can be shown by direct inspection that they satisfy the relations
\begin{eqnarray}
	\frac{\partial f_1}{\partial r}+\frac{\partial f_2}{\partial p_r}=0,\\
	\frac{\partial g_{1m}}{\partial r}+\frac{\partial g_{2m}}{\partial p_r}=0,\\
	\frac{\partial g_{1s}}{\partial r}+\frac{\partial g_{2s}}{\partial p_r}=0,
\end{eqnarray}
which imply that the dynamical system is Hamiltonian in the canonical conjugate variables $r$ and $p_r$.

\subsection{Galactic spacetime}

The metric tensor describing the geometry of the spacetime of a black hole surrounded by a dark matter halo proposed by~\cite{cardosoEtal2022PRD}  is of the form of Eq.~(\ref{genmetric}) and assumes an anisotropic fluid for the halo, with only  tangential pressure (vanishing radial pressure). The mass function $\mu(r)$, which is related to $g(r)$ in Eq.~(\ref{genmetric}) by $g(r)=\left[1-2\mu(r)/r\right]^{-1}$, is given by
\begin{equation}
	\mu(r)=\frac{r_0}{2}+\frac{Mr^2}{\left(a+r\right)^2}\left(1-\frac{r_0}{r}\right)^2
\label{mcardoso}
\end{equation}
and approximates the Schwarzschild metric of a black hole of mass $r_0/2$ at small distances. At large distances, this metric corresponds to a Hernquist dark matter profile for a halo of mass $M$ and typical length scale $a$. The function $f(r)$ in Eq.~(\ref{genmetric}) can be determined by Einstein field equations and reads
\begin{eqnarray}
	f(r)&=&\left(1-\frac{r_0}{r}\right)e^\Upsilon,\\
	\Upsilon&=&-\pi\sqrt{\frac{M}{\xi}}+2\sqrt{\frac{M}{\xi}}\arctan\frac{r+a-M}{\sqrt{M\xi}},\\
	\xi&=&2a-M+2r_0.
\label{fcardoso}
\end{eqnarray}

A generalization of this method, proposed by \cite{konoplyaZhidenko2022ApJ}, matches the geometry of a black hole with that of any member of a family of dark matter halo profiles, in a self-consistent manner. This includes the Hernquist profile. Although the functions $f(r)$ and $g(r)$ in the metric are to be numerically obtained, an analytical approximation is offered for the Hernquist profile, which matches quite well Eqs.~(\ref{mcardoso})--(\ref{fcardoso}):
\begin{eqnarray}
	f(r)&=&\left(1-\frac{r_0}{r}\right)\left(1-\frac{2M}{r+a}\right),\label{fkanoplya}\\
	\mu(r)&=&\frac{r_0}{2}+\frac{Mr^2}{\left(r+a\right)^2}\left(1-\frac{r_0}{r}\right).\label{mkanoplya}
\end{eqnarray}
In our calculations we consider the spacetime described by Eqs. (\ref{fkanoplya})--(\ref{mkanoplya}).

\subsection{Quadrupolar oscilations}

We now go back to the case of an extended body. In order to model a pulsating star, we consider a time-dependent profile for $j_m$ with angular frequency $\Omega$,
\begin{equation}
	j_m(\tau)=q_0\left[1+\sin(\Omega\tau)\right],
\label{jm}
\end{equation}
where $q_0$ is a small positive parameter. Being a mass quadrupole, $j_m$ remains always nonnegative, effectively describing the fundamental mode of linear radial pulsations \cite{carroll2006introduction}. We make the simplifying assumption $j_s(\tau)=0$, which physically corresponds to the case where the mass quadrupoles are dominant over the stress quadrupoles. However, we remark that our formalism also deals with nontrivial configurations for the stress quadrupole, as can be seen from Eqs.~(\ref{rdot}) and~(\ref{prdot}).

\section{Point-particle homoclinic orbits}
\label{pointparticle}

The motion of a point particle in the spacetimes here considered may be obtained from Eqs.~(\ref{rdot}) and~(\ref{prdot}) in the limit when both $j_s$ and $j_m$ vanish, in which case the orbit is a geodesic. From the expression
\begin{equation}
	p_\mu p^\mu=-m^2,
\end{equation}
we can interpret $m$ as the mass of the extended body, which is generally not constant in (proper) time; it clearly becomes constant in the point-particle limit, $m\equiv m_0$. The equation above may be solved for $p_r$ and then, substituting it on Eqs.~(\ref{rdot}) and~(\ref{prdot}) and setting $j_m=j_s=0$, we obtain
\begin{equation}
	\frac{\dot{r}^2}{2}+V=0,
\label{rp2}
\end{equation}
where we were led to define the potential
\begin{equation}
	V=\frac{1}{2g(r)}+\frac{\ell^2}{2r^2g(r)}-\frac{e^2}{2f(r)\,g(r)},
\label{veff}
\end{equation}
with $e=E/m_0$ and $\ell=L/m_0$. The local minima and maxima of $V$ determine the position of the stable and unstable circular orbits, respectively, for a given value of $e$ and $\ell$. Also, a circular orbit requires $\dot{r}=0$. Hence, the conditions for a circular orbit of radius $r_c$ can be cast as
\begin{eqnarray}
	V(r_c)&=&0\\
	V'(r_c)&=&0.
\end{eqnarray}

One may solve this set of equations for the orbit parameters $\ell$ and $e$. It follows that any unstable circular orbit of radius $r=r_{un}$ is characterized by the values of the parameters
\begin{eqnarray}
	\ell=\sqrt{\frac{{r_{un}}^3f'(r_{un})}{2f(r_{un})-r_{un}f'(r_{un})}},\label{lun}\\
	e=\frac{f(r_{un})}{\sqrt{f(r_{un})-\frac{1}{2}r_{un}f'(r_{un})}},\label{eun}
\end{eqnarray}
along with the usual condition $V''(r_{un})<0$. This last condition will restrict the possible values of $r_{un}$ in a manner that depends on the exact form of the function $f(r)$. It is possible then to parametrize the potential $V$ by $r_{un}$ by substituting Eqs.~(\ref{lun}) and~(\ref{eun}) in Eq.~(\ref{veff}). An explicit expression for $V$ is given in Eq.~(\ref{vexpl}), for the metric~(\ref{fkanoplya})--(\ref{mkanoplya}).

\begin{figure}
	\includegraphics[width=\columnwidth]{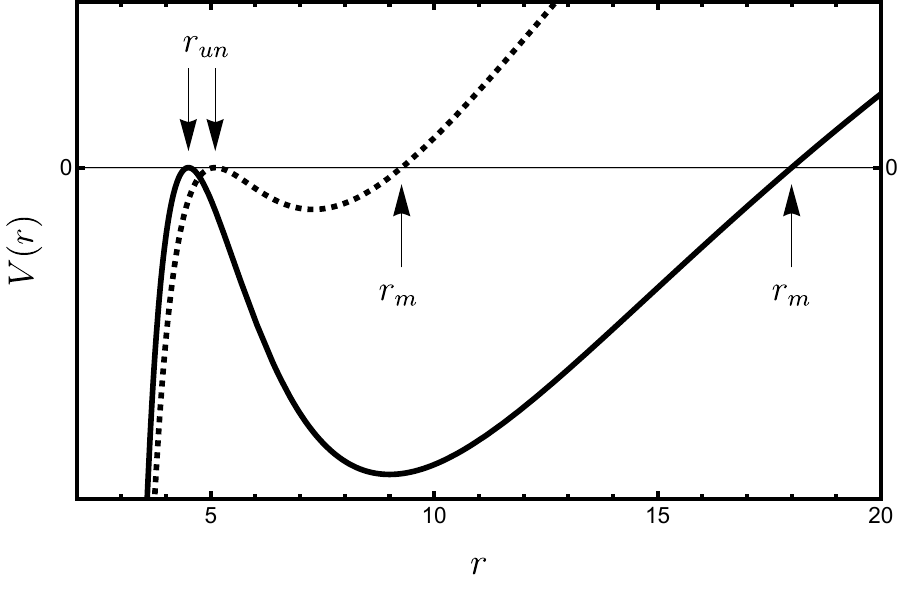}
	\caption{Potential $V(r)$ for $r_{un}=4.5$ (continuous line) and $r_{un}=5.1$ (dashed line). The arrows indicate the minimum and maximum radii for the homoclinic orbit. Units are given by~(\ref{r0})--(\ref{agal}).}
	\label{graphveff}
\end{figure}

\begin{figure}
	\includegraphics[width=\columnwidth]{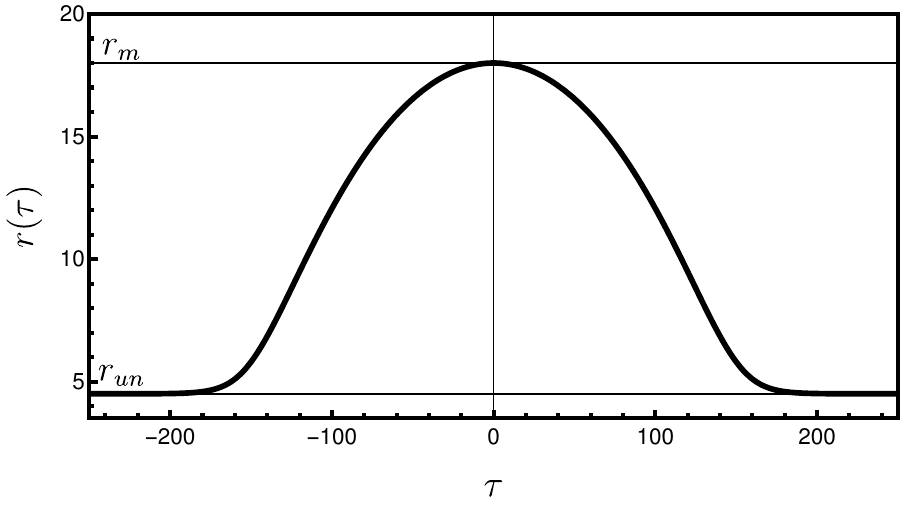}
	\includegraphics[width=\columnwidth]{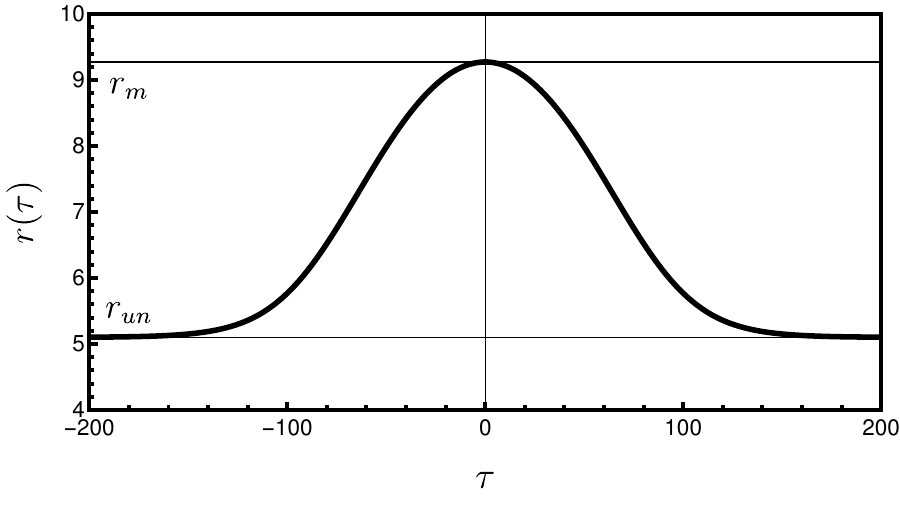}
	\caption{Plot of $r(\tau)$ for the homoclinic orbits characterized by $r_{un}=4.5$ (top) and $r_{un}=5.1$ (bottom).}
	\label{graphrtau}
\end{figure}

For the metrics considered here, contrarily to what happens in Schwarzschild spacetime \cite{mosnaRodriguesVieira2022PRD}, the point-particle homoclinic orbits cannot be obtained analytically, only numerically. We adopt henceforth the following values for the parameters in the metric of Eqs.~(\ref{fkanoplya})--(\ref{mkanoplya}):
\begin{eqnarray}
	r_0=2,\label{r0}\\
	M=10,\label{Mgal}\\
	a=10^5,\label{agal}
\end{eqnarray}
in units of the mass of the black hole, $M_{BH}$. With these values, the range for the parameter $r_{un}$ corresponding to bounded orbits differs only slightly from the same range for the Schwarzschild metric, that is, $4<r_{un}<6$. The turning point, $r_m$, may also be calculated. Figure~\ref{graphveff} shows the potential $V$ for $r_{un}=4.5$ and $r_{un}=5.1$, with the respective turning points, $r_m=18.00$ and $r_m=9.27$. Figure~\ref{graphrtau} shows, for each of these potentials, the orbits that reach the returning point in $\tau=0$ and tend asymptotically to the unstable radius $r_{un}$ (the so-called homoclinic orbits associated with $r_{un}$). In what follows, we restrict ourselves to presenting the results only for the metric in Eqs.~(\ref{fkanoplya}) and~(\ref{mkanoplya}), with the values of the parameters mentioned above, since the others metrics give us very similar results.

\section{Homoclinic Chaos}
\label{chaos}

A homoclinic orbit in an autonomous two-dimensional dynamical system, like the ones just obtained in Sec.~\ref{pointparticle}, may have its integrability broken by a time-periodic perturbation. Melnikov's method, which we briefly recall now, allows the identification of this phenomenon by the simple evaluation of an integral.

An integrable Hamiltonian dynamical system with canonical variables $(r,p_r)$, when subject to a time-periodic perturbation of (angular) frequency $\Omega$, comes to obey the canonical equations
\begin{eqnarray}
	\dot{r}=f_1(r,p_r)+\epsilon\lambda_1(r,p_r,\tau)\label{drmel}\\
	\dot{p_r}=f_2(r,p_r)+\epsilon\lambda_2(r,p_r,\tau)\label{dprmel}
\end{eqnarray}
with a small parameter $\epsilon$ and with $\lambda_i$ periodic in $\tau$, with period $2\pi/\Omega$. If the unperturbed system presents an unstable fixed point, in our case $(r=r_{un},p_r=0)$, then the perturbation may split its homoclinic orbit in a stable and an unstable manifolds of the stroboscopic map (defined by $\tau_n = 2n\pi/\Omega$, for every integer $n$). The transverse crossings of these manifolds give rise to a homoclinic tangle and thus to a chaotic behavior in the system. The distance between the stable and unstable manifolds at a particular time $\tau_0$ along the direction perpendicular to the unperturbed orbit at the same time is proportional, in first order in $\epsilon$, to Melnikov's integral \cite{holmes1990PhysRep, lichtenbergLieberman1992}
\begin{equation}
	M(\tau_0)=\int_{-\infty}^{\infty}(f_1\lambda_2-f_2\lambda_1)(r(\tau),p_r(\tau),\tau+\tau_0)\,d\tau,
\label{melint}
\end{equation}
where the integrand is evaluated at the unperturbed orbit. Therefore, isolated zeros of $M(\tau_0)$ represent the tranversal intersections of the manifolds, whereas $M(\tau_0)$ being identically zero indicates the preservation of the homoclinic loop and the absence of chaos.

The comparison of Eqs.~(\ref{drmel})--(\ref{dprmel}) with Eqs.~(\ref{rdot})--(\ref{prdot}) shows that in our case $\epsilon\lambda_1=j_m(\tau)g_{1m}$ and $\epsilon\lambda_2=j_m(\tau)g_{2m}$, with $j_m$ given by Eq.~(\ref{jm}) and $f_1$, $f_2$, $g_{1m}$, and $g_{2m}$ given by Eqs.~(\ref{f1})--(\ref{g2s}). By the symmetry of the homoclinic orbit with respect to instant $\tau=0$, when $r=r_m$, Melnikov's integral can be written as
\begin{equation}
	M(\tau_0)=2\cos(\Omega\,\tau_0)\,K(\Omega),
\label{melint2}
\end{equation}
which singles out the dependence on $\tau_0$ and where
\begin{equation}
	K(\Omega)=\int_{r_{un}}^{r_m}(f_1g_{2m}-f_2g_{1m})(r)\sin(\Omega\,\tau(r))\frac{d\tau}{dr}\,dr.
\end{equation}
In this equation, $d\tau/dr$ is evaluated along the unperturbed homoclinic orbit and can be obtained from Eq.~(\ref{rp2}).

\begin{figure}
	\includegraphics[width=\columnwidth]{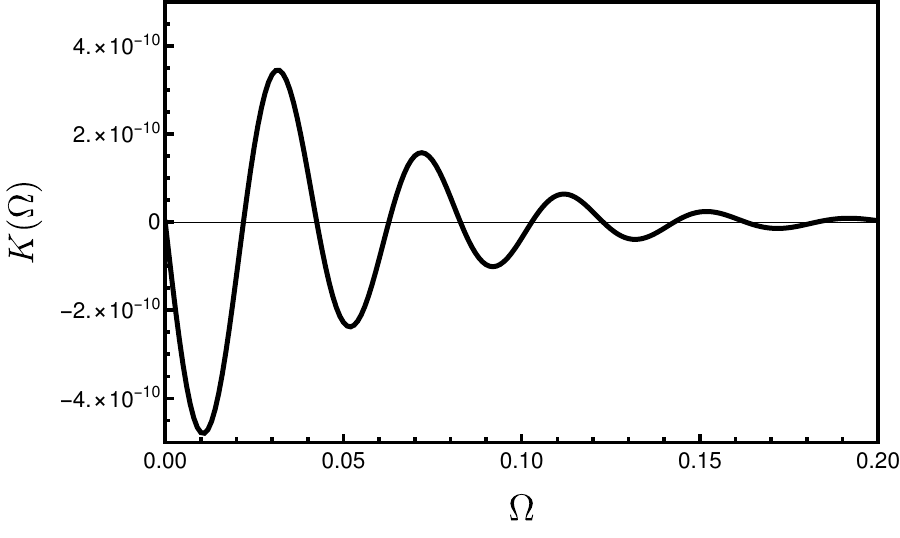}
	\includegraphics[width=\columnwidth]{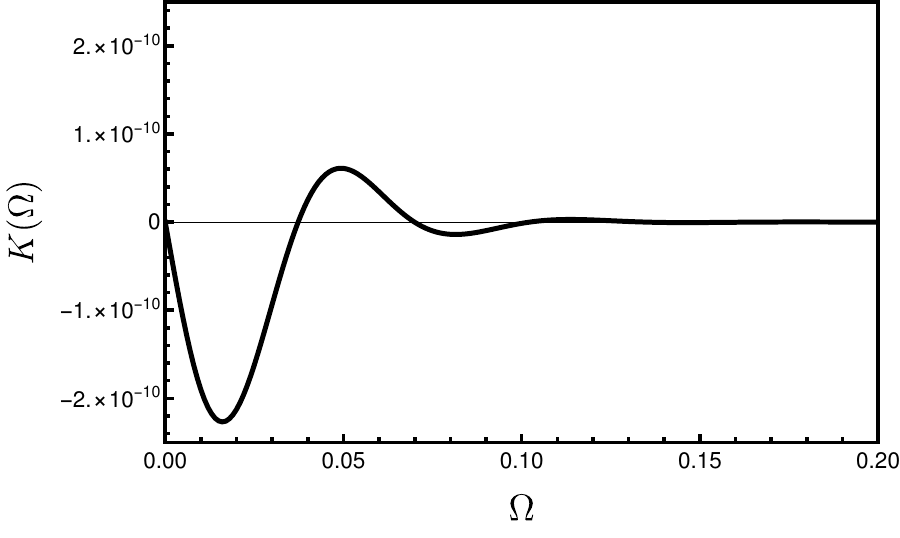}
	\caption{Plot of the function $K(\Omega)$ for $r_{un}=4.5$ (top) and $r_{un}=5.1$ (bottom). The oscillatory behavior ensures that $K\neq0$ at almost every value of $\Omega$.}
	\label{graphkomega}
\end{figure}

Figure~\ref{graphkomega} shows the behavior of $K$ as a function of $\Omega$, for $r_{un}=4.5$ and $r_{un}=5.1$. It can be seen that $K=0$ only for isolated values of $\Omega$. For all the values of $\Omega$ for which $K\neq0$, we can see from Eq.~(\ref{melint2}) that Melnikov's integral has a countable infinite number of simple zeros, corresponding to a countable infinite number of intersections between the unstable and stable manifolds associated with the (perturbed) unstable fixed point, a signature of homoclinic chaos in that region. We stress that, in practice, this chaotic behavior has a transient nature, since the body eventually falls into the black hole in a finite proper time (see \cite{mosnaRodriguesVieira2022PRD} for more details).

\section{Astrophysical implications}
\label{implications}

Since Melnikov's method detects homoclinic chaos near unstable orbits for any value of the perturbation parameter, and since the perturbation is related to the body's size ($q_0 \sim m R^2$ to within a factor of one, where $R$ is the characteristic radius of the body), the approach presented here can be applied to any small spherical test body around a black hole, once its quadrupole parameters oscillate in (proper) time. In particular, pulsating stars around supermassive black holes, if sufficiently close, would acquire chaotic behavior due to their oscillating mass distribution, no matter how small this oscillation is.

Also, although the model presents an oscillation which varies between a point particle and a sphere, it can be shown that the same formalism applies to small oscillations around a finite-size sphere, such as $j_m = q_0 (1 + \varepsilon \sin(\Omega \tau))$, with $\varepsilon$ small, yielding the same qualitative results regarding chaotic orbital motion. This model of linear pulsation is the simplest prescription for modeling regular variable stars \cite{carroll2006introduction}, corresponding to the fundamental mode of the Fourier expansion of a general periodic brightness (and density) variation.

Therefore, size does not matter: any star around a supermassive black hole is subject to this effect. Typical values of $q_0$ range from $10^{-12}$ to $10^{-5}$, corresponding to stellar masses between 0.1 and 100 solar masses.

However, the pulsation frequency $\Omega$ must be of the same order as the other frequencies in the system in order for this chaotic behavior to be manifest. If $\Omega$ is much larger than the other characteristic frequencies, then there will be no time for the system to “feel” the perturbation in its orbital dynamics. On the other hand, if $\Omega$ is much smaller than the characteristic frequencies of the system, then the time for chaos to be noticeable will be much larger than the typical period of orbital revolution. The angular frequency of a circular orbit in the unstable region, $r \sim \alpha M_{\rm BH}$, is given by
\begin{equation}\label{eq:omegaphi}
	\omega_\phi \approx \sqrt{\frac{GM_{\rm BH}}{r^3}} = \frac{c^3}{G M_{\odot}}\,\alpha^{-3/2} \,\frac{1}{x},
\end{equation}
where $M_{\rm BH} = x M_{\odot}$ and $\alpha$ is dimensionless. For $\alpha = 4$, corresponding to the region inside the marginally stable orbit, $\omega_\phi \approx 10^{-2}$\,Hz for a $10^6 M_\odot$ black hole, with an azimuthal period of $T_\phi \approx 4$\,min. For a supermassive black hole of $\sim 10^9 M_\odot$, $\omega_\phi \approx 10^{-5}$\,Hz, corresponding to $T_\phi \approx 70$\,h. Thus, for black holes in the mass range $10^6$--$10^{10} M_\odot$, the period of revolution ranges from minutes to hours, lying within the range of brightness variations of oscillating stars \cite{carroll2006introduction}.

We now compare the innermost stable circular orbit (ISCO) radius with the tidal disruption radius $R_T$, given by
\begin{equation} \label{eq:RT}
R_T \approx \frac{R_* (M_{\rm BH}/M_*)^{1/3}}{\beta},
\end{equation}
where $R_*$ and $M_*$ are the star's radius and mass, respectively, and
$\beta$ reflects the internal structure of the star, roughly determined by the ratio of its central to mean density, and varies from 1 to 7 depending on the equation of state \cite{coughlinNixon2022MNRASL}. Since $R_{ISCO} \approx 6\, GM_{\rm BH}/c^2$, this yields
\begin{equation}\label{eq:RTRisco}
\frac{R_T}{R_{ISCO}} \approx \frac{0.1\, c^2 G^{-1} R_* M_*^{-1/3} M_{\rm BH}^{-2/3}}{\beta}.
\end{equation}
For a Sun-like star ($M_* = 1 M_\odot$ and $R_* \approx 7 \times 10^5$\,km), 
\begin{equation}\label{eq:RTsun}
\frac{R_T}{R_{ISCO}} \approx \frac{5 \times 10^4 \left( \frac{M_{\rm BH}}{M_\odot} \right)^{-2/3}}{\beta}.
\end{equation}
Thus, for supermassive black holes with $M_{\rm BH} > 5 \times 10^5 M_\odot$, corresponding to $\beta = 7$, such stars may enter the ISCO without being tidally disrupted. A more conservative estimate, for $\beta$=2, gives a lower value $M_{\rm BH} > 3.6 \times 10^6 M_\odot$ for tidal disruption after entering the ISCO. In any case, these ranges include the supermassive black hole SgrA* in the center of our Galaxy, with a mass of $\sim 4 \times 10^6 M_\odot$, and most other central black holes in galaxies.

Still regarding possible observable effects, the proper frequencies of radiation emitted by the star will be redshifted, when observed at ``infinity,'' according to the star's position and its line-of-sight speed. Therefore, since the time series of $r(t)$ and $\vec{v}_{\rm{los}}(t)$ are chaotic, the time series of the characteristic frequencies of the star will be erratic in terms of coordinate time.
Also, since $dt = F(r, p_r, ....) d\tau$ depends on the orbital coordinates, the period of pulsation as seen at infinity (by integrating over $2\pi/\Omega$ in $d\tau$) will depend on the star's position and velocity along its orbit, and since chaotic motion gives large wanderings to the star over very small (proper) time spans, the observed ``period'' of pulsation at infinity will also appear erratic in the star's light curve when measured over a long time. These effects are, in principle, present in the measurements; however, their ``amplitude'' or significance will depend on the system in question and on the particular orbit of the star. By the same arguments as above, imprints of this chaoticity may appear in the  gravitational-wave signals of extreme mass-ratio inspirals, detected on Earth by forthcoming observatories such as the Laser Inteferometer Space Antenna~\cite{lisa2023LRR}. 

In summary, the similarity between the orders of magnitude of the azimuthal and of the brightness oscillation periods, and the possibility of stars approaching the ISCO without being tidally disrupted imply that this effect could be in principle observable in our Galaxy, given the mass of SgrA*, and potentially in other galaxies with more massive black holes, such as M87, though current resolution is not precise enough to track individual stars in other galaxies.

There is also the case of solar-mass BH + NS binaries, for which neutron stars mostly do not suffer tidal disruption before plunging into the BH \cite{foucart2020FrASS}. We then may have this chaotic imprint on the time series of waveforms of BH + NS mergers if the NS pulsates, and also on the corresponding electromagnetic signal. However the present model would only be an idealization, since we would assume no distortion from spherical symmetry (before the NS fluid becomes sufficiently distorted) and also that the NS would behave as a test body, which evidently it is not since the masses of the two objects are comparable in general.
Also, for a pulsating test body around a stellar-mass BH, its size should be small enough so that the parameter $q_0$ is small compared to the BH mass since  we are in the test-fluid approximation.
On the other hand, we must note that the present framework opens the possibility of studying this kind of phenomenon in more realistic scenarios.

\section{Conclusion}
\label{conclusion}

It is important to note that there is a remarkable qualitative difference between orbits of extended bodies in vacuum and nonvacuum black-hole spacetimes. Although periodically shape-changing bodies may present orbital chaos in (pure) Schwarzschild spacetime, this does not happen if the pulsating body is spherically symmetric \cite{mosnaRodriguesVieira2022PRD}. In this way, the onset of chaos in the orbital dynamics of spherically symmetric pulsating stars, as presented here, is a direct consequence of the presence of ambient gravitating matter around the (Schwarzschild) black hole, what would also be expected to happen in the many other exact solutions of this kind existent in the literature, e.g.,  \cite{cardosoEtal2022PRD, konoplyaZhidenko2022ApJ, xuEtal2018JCAP, nampalliwarEtal2021ApJ, jusufi2023EPJC, shenEtal2024PhLB, datta2024PhRvD, daghigKunstatter2024PhRvD,	yangEtal2024EPJC}, among others.

In our context, this means that the most common type of stellar pulsation (corresponding to the fundamental radial mode of oscillation in the linear regime) does not alter its translational motion around vacuum Schwarzschild black holes, which remains regular. These stars will present chaos in their orbits only in the presence of surrounding matter, which in our case was treated as a gravitating dark matter halo based on previous ``superposed'' solutions existing in the literature.
We stress, however, that any ambient matter would lead to this effect. In particular, for supermassive black holes in the center of galaxies (including our own), the galactic bulge, generally present in the vicinity of the black hole, will also play this role when treated as self gravitating. Therefore even in the absence of dark matter, the central structures encompassing galactic black holes will guarantee chaotic behavior of these stars. Also, since chaos is manifest for sinusoidal quadrupole oscillations, more complicated radial and nonradial pulsations of the star will also lead to chaotic orbital motion. As mentioned in the text, this chaotic behavior may leave imprints in light curves of the stellar brightness and/or gravitational wave signals (from the star's infall into the black hole) detected on Earth.

As a final remark, from a purely theoretical point of view, the phenomenon that a spherical variable star requires a matter background to have its orbit altered by its own internal motion (in the quadrupole approximation) is interesting in its own right, generalizing the corresponding results in Newtonian gravity \cite{vieiraMosna2022CSF}.

\section*{ACKNOWLEDGMENTS}
R.A.M. was partially supported by Conselho Nacional de Desenvolvimento Cient\'{\i}fico e Tecnol\'{o}gico (CNPq, Brazil) under Grant No. 316780/2023-5.

\section*{DATA AVAILABILITY}
No data were created or analyzed in this study.




\appendix
\section{EQUATIONS}
\label{appendix}

In Dixon's formalism, the force $F_\mu$ up to quadrupolar order is given by Eq.~(\ref{forcedixon}). 

We computed it following the procedure in~\cite{mosnaRodriguesVieira2022PRD}, using a moving reference frame that simplifies the components of the quadrupole moment tensor. The resulting nonzero components of the force, $F_t$, $F_r$, $F_\phi$ are presented in the equations below, where $A^{-1}=3 f(r)\,g(r)^2(f(r) (g(r)\,p_{\phi }^2+r^2 p_r^2)-r^2 g(r)\, p_t^2)$.

After the procedure outlined in the main text, we obtain the functions $f_1(r,p_r)$, $g_{1m}(r,p_r)$, $g_{1s}(r,p_r)$, $f_2(r,p_r)$, $g_{2m}(r,p_r)$, and $g_{2s}(r,p_r)$ that appear in Eqs.~(\ref{rdot})--(\ref{prdot}). These functions are displayed below, with $B^{-1}=E^2 r^2 g(r)-f(r)(L^2 g(r)+r^2 p_r^2)$. In these equations, we have renamed the parameters $p_t$ and $p_\phi$ to $-E$ and $L$, respectively.

Finally, the potential defined in Eq.~(\ref{veff}) can be parametrized by the radius of the unstable circular orbit, $r_{un}$, through Eqs.~(\ref{lun})--(\ref{eun}). The explicit form for $V$, when the metric is determined through Eqs.~(\ref{fkanoplya})--(\ref{mkanoplya}), is shown in the last equation.

\begin{widetext}

\begin{align}
F_t=&\,2 A (j_m+j_s) r p_r p_t f'(r) (g(r) f'(r)+f(r) g'(r)),&
\label{ft}
\\
	F_r=&\,\frac{A}{2r^3f(r)^2g(r)}(2 g(r)^3 j_m p_t^2 f'(r)^3 r^5+2 f(r) g(r)^2 f'(r) (f'(r) (j_m (p_t^2 g'(r)-p_r^2 f'(r)) r^2+g(r) (j_s p_{\phi }^2 f'(r)\nonumber\\
	&-2 r j_m p_t^2))-2 r^2 g(r) j_m p_t^2 f''(r)) r^3+f(r)^2 g(r) (2 j_m f'(r) g'(r) (p_t^2 g'(r)-p_r^2 f'(r)) r^3+g(r) (-3 r^2 j_m\times\nonumber\\
	&g'(r) f''(r) p_t^2+2 j_s f'(r)^2 (p_{\phi }^2 g'(r)-2 r p_r^2)-r j_m f'(r) (p_t^2 (4 g'(r)+r g''(r))-4 r p_r^2 f''(r))) r+2 g(r)^2\times\nonumber\\
	&(r^2 j_m(2 f''(r)+r f^{(3)}(r)) p_t^2+(j_m-j_s) p_{\phi }^2 f'(r)^2-2 r f'(r) (j_m p_t^2+j_s p_{\phi }^2 f''(r)))) r^2+f(r)^3 (-2 j_m p_r^2\times\nonumber\\
	&f'(r) g'(r)^2 r^4+g(r) (r^2 j_m f'(r) g''(r) p_r^2+g'(r) (2 j_s f'(r) (p_{\phi }^2 g'(r)-2 r p_r^2)+r (3 r j_m p_r^2 f''(r)\nonumber\\
	&-8 j_s p_t^2 g'(r)))) r^2-8 g(r)^4 j_s p_t^2 r-g(r)^2 (f'(r) (4 r j_s p_r^2+p_{\phi }^2 (2 (j_s-j_m) g'(r)+r j_s g''(r)))+r (j_s (3 p_{\phi }^2\times\nonumber\\
	&g'(r)-4 r p_r^2) f''(r)+2 r (r j_m p_r^2 f^{(3)}(r)-2 j_s p_t^2 g''(r)))) r+2 g(r)^3 (4 r j_s p_t^2+p_{\phi }^2 (j_s f^{(3)}(r) r^2+(j_m-j_s)\times\nonumber\\
	&(f'(r)-r f''(r))))) r+2 f(r)^4 (-4 j_m p_r^2 g'(r)^2 r^4+2 g(r) (r^2 j_m g''(r) p_r^2+g'(r) ((j_s-j_m) p_{\phi }^2 g'(r)\nonumber\\
	&-r (j_m+j_s) p_r^2)) r^2+g(r)^2(p_{\phi }^2 ((j_m+j_s) g'(r)+r (j_m-j_s) g''(r))-4 r j_s p_r^2) r-4 g(r)^4 j_m p_{\phi }^2\nonumber\\
	&+4 g(r)^3 (r^2 j_s p_r^2+j_m p_{\phi }^2))),\label{fr}
\\
	F_\phi=&\,\frac{A}{r}(p_r p_{\phi } (j_m+j_s) (r^2 g(r) f'(r)^2+r f(r) (f'(r) (r g'(r)+2 g(r))-2 r g(r) f''(r))+f(r)^2 (2 r g'(r)& \nonumber\\
	&+4 (1-g(r)) g(r)))),\label{fphi}
\\
	f_1=&\,r f(r) p_r \sqrt{\frac{B}{f(r) g(r)}},&\label{f1}
\\
	g_{1m}=&\,(3 g(r))^{-1}(B^2 p_r (r^2 g(r) f'(r) (4 E^2 r-L^2 f'(r))+r f(r) (r g'(r) (4 E^2 r-L^2 f'(r))-2 L^2 g(r) (f'(r)\nonumber\\
	&-r f''(r)))+2 L^2 f(r)^2 (2 (g(r)-1) g(r)-r g'(r)))),\label{g1m}
\\
	g_{1s}=&\,g_{1m},&\label{g1s}
\\
	f_2=&-\frac{\sqrt{B} (E^2 r^3 g(r)^2 f'(r)-f(r)^2 (r^3 p_r^2 g'(r)+2 L^2 g(r)^2))}{2 r^2 (f(r) g(r))^{3/2}},&\label{f2}
\\
	g_{2m}=&\,(6 r^3 f(r)^3 g(r)^3)^{-1}(B^2 (-2 E^4 g(r)^4 f'(r)^3 r^7+E^2 f(r) g(r)^3 f'(r) (4 E^2 g(r) f''(r) r^2+f'(r) (2 (2 p_r^2 f'(r)\nonumber\\
	&-E^2 g'(r)) r^2+g(r) (4 r E^2+3 L^2 f'(r)))) r^5-f(r)^2 g(r)^2 (2 f'(r) (p_r^2 f'(r)-E^2 g'(r)){}^2 r^3+g(r) (-3 r^2\times\nonumber\\
	&g'(r) f''(r) E^4+f'(r)^2 (8 r p_r^2-3 L^2 g'(r)) E^2-r f'(r) (E^2 (4 g'(r)+r g''(r))-8 r p_r^2 f''(r)) E^2+2 L^2 p_r^2\times\nonumber\\
	&f'(r)^3) r+2 E^2 g(r)^2 (E^2 (2 f''(r)+r f^{(3)}(r)) r^2+f'(r) (3 L^2 f''(r)-2 E^2) r+5 L^2 f'(r)^2)) r^4+f(r)^3 g(r)\times\nonumber\\
	&(-2 p_r^2 f'(r)g'(r) (p_r^2 f'(r)-2 E^2 g'(r)) r^5+g(r) (-3 L^2 f'(r)^2 g'(r) p_r^2-6 E^2 r^2 g'(r) f''(r) p_r^2+2 f'(r) (-2 E^2\times\nonumber\\
	&r g'(r) p_r^2+r^2 (2 p_r^2 f''(r)-E^2 g''(r)) p_r^2+E^2 L^2 g'(r)^2)) r^3+g(r)^2 (r (4 r^2 f^{(3)}(r) p_r^2+(4 r p_r^2-3 L^2 g'(r))\times\nonumber\\
	&f''(r)) E^2+4 L^2 p_r^2 f'(r)^2-f'(r) ((E^2 (10 g'(r)+r g''(r))-4 r p_r^2 f''(r)) L^2+4 E^2 r p_r^2)) r^2\nonumber\\
	&-4 E^2 L^2 g(r)^4 f'(r) r+2 L^2 g(r)^3 (r f'(r) E^2+r^2 (5 f''(r)+r f^{(3)}(r)) E^2+L^2 f'(r)^2)) r^2+f(r)^4\times\nonumber\\
	&(-2 p_r^4 f'(r) g'(r)^2 r^6+g(r) p_r^2(r^2 f'(r) g''(r) p_r^2+3 g'(r) (r^2 f''(r) p_r^2+(4 E^2 r-L^2 f'(r)) g'(r))) r^4+g(r)^2\times\nonumber\\
	&(-2 r^3 f^{(3)}(r) p_r^4+g'(r) (4 r E^2+L^2 (2 f'(r)+5 r f''(r))) p_r^2+r (L^2 f'(r)-4 E^2 r) g''(r) p_r^2+4 E^2 L^2 g'(r)^2) r^3\nonumber\\
	&+16 E^2 L^2 g(r)^5 r+2 L^2 g(r)^3(f'(r) (g'(r) L^2+3 r p_r^2)+r (E^2 g'(r)-r (g''(r) E^2+p_r^2 (3 f''(r)+r f^{(3)}(r))))) r\nonumber\\
	&-2 g(r)^4 ((r f''(r)-f'(r)) L^4+8 E^2 r L^2)) r+2 f(r)^5 (-4 p_r^4 g'(r)^2 r^6+g(r) p_r^2 (-2 r g'(r) p_r^2+2 r^2 g''(r) p_r^2\nonumber\\
	&-7 L^2 g'(r)^2) r^4-L^2 g(r)^2 (r g'(r) p_r^2-3 r^2 g''(r) p_r^2+2 L^2 g'(r)^2) r^2+L^2 g(r)^3 (r g''(r) L^2+8 r p_r^2+(L^2\nonumber\\
	&+2 r^2 p_r^2) g'(r)) r-4 L^4 g(r)^5+4 g(r)^4 (L^4-2 L^2 r^2 p_r^2)))),\label{g2m}
\\
	g_{2s}=&\,(6 r^2 f(r)^2 g(r)^2)^{-1}(B^2 (-E^2 L^2 g(r)^3 f'(r)^3 r^4+L^2 f(r) g(r)^2 f'(r) (2 E^2 g(r) f''(r) r^2+f'(r) ((2 p_r^2 f'(r)\nonumber\\
	&-E^2 g'(r)) r^2+g(r) (2 L^2 f'(r)-2 E^2 r))) r^2+f(r)^2 g(r) (-((p_r^2 f'(r)-2 E^2 g'(r)) (4 r g'(r) E^2\nonumber\\
	&+f'(r) (4 r p_r^2-L^2 g'(r))) r^3)+4 E^2 g(r)^3 (2 E^2 r-L^2 f'(r)) r+g(r) (E^2 ((3 L^2 g'(r)-4 r p_r^2) f''(r)\nonumber\\
	&-4 E^2 r g''(r)) r^2+f'(r) ((E^2 (r g''(r)-2 g'(r))-4 r p_r^2 f''(r)) L^2+4 E^2 r p_r^2) r+f'(r)^2 (2 L^4 g'(r)\nonumber\\
	&-4 L^2 r p_r^2)) r-2 g(r)^2 (f'(r)^2 L^4+r f'(r) (2 L^2 f''(r)-5 E^2) L^2+E^2 r^2 (4 E^2+L^2 (r f^{(3)}(r)-f''(r))))) r\nonumber\\
	&-2 f(r)^4 (p_r^2 g'(r) (2 r p_r^2-L^2 g'(r)) r^3+g(r) (4 r^2 p_r^4+L^2 (r g'(r) p_r^2+r^2 g''(r) p_r^2-2 L^2 g'(r)^2)) r\nonumber\\
	&-g(r)^2 (-r g''(r) L^4+(L^2+2 r^2 p_r^2) g'(r) L^2+4 r^3 p_r^4))+f(r)^3 (p_r^2 g'(r) (f'(r) (L^2 g'(r)-4 r p_r^2)\nonumber\\
	&-4 E^2 r g'(r)) r^4+g(r) (r (r g'(r) (4 E^2-L^2 f''(r)) p_r^2+4 r^2 (g''(r) E^2+p_r^2 f''(r)) p_r^2-12 E^2 L^2 g'(r)^2)\nonumber\\
	&-f'(r) (-2 g'(r)^2 L^4+6 r p_r^2 g'(r) L^2+r^2 p_r^2 (g''(r) L^2+4 p_r^2))) r^2+g(r)^2 ((r (6 r g''(r) E^2+g'(r) (2 E^2\nonumber\\
	&-3 L^2 f''(r))+2 r p_r^2 (f''(r)+r f^{(3)}(r)))-f'(r) ((2 g'(r)+r g''(r)) L^2+2 r p_r^2)) L^2+16 E^2 r^2 p_r^2) r\nonumber\\
	&+2 g(r)^3 (L^4 (r (f''(r)+r f^{(3)}(r))-f'(r))-8 E^2 r^3 p_r^2)))),\label{g2s}
\\
	V=&-[(a^2+2 a r+r (r-2 M)) ((r-r_0) (r-r_{\text{un}}) r_{\text{un}} (r_{\text{un}}+r) (a-2 M+r) (2 M (r_0-r_{\text{un}}) r_{\text{un}}\nonumber\\
	&-r_0 (a+r_{\text{un}}) (a-2 M+r_{\text{un}}))-2 r^2 (r_0-r_{\text{un}}) (a-2 M+r_{\text{un}}) ((r-r_0) r_{\text{un}} (a-2 M+r) (a+r_{\text{un}})\nonumber\\
	&+r (a+r) (r_0-r_{\text{un}}) (a-2 M+r_{\text{un}})))][2 r^3 (a+r)^2 r_{\text{un}} (a-2 M+r) (r_0 ((6 a-8 M) r_{\text{un}}+3 a (a-2 M)\nonumber\\
	&+3 r_{\text{un}}^2)-2 r_{\text{un}} ((2 a-3 M) r_{\text{un}}+a (a-2 M)+r_{\text{un}}^2))]^{-1}.&\label{vexpl}
\end{align}

\end{widetext}

\end{document}